\documentclass[prl,aps,twocolumn,superscriptaddress,showpacs,floatfix]{revtex4}
\usepackage{epsfig,graphicx,latexsym}
\begin{document}

\title{Thermal Casimir effect between random layered dielectrics}

\author{D. S. Dean}
\address{Kavli Institute for Theoretical Physics, University of California Santa Barbara, CA 93106, U.S.A.}
\address{Laboratoire de Physique Th\'eorique, IRSAMC, Universite Paul Sabatier, 118 Route de Narbonne, 31062 Toulouse Cedex 4, France}
\author{R.R. Horgan}
\address{Kavli Institute for Theoretical Physics, University of California Santa Barbara, CA 93106, U.S.A.}
\address{DAMTP, CMS, University of Cambridge, Cambridge,  CB3 0WA, UK. \\}
\author{A. Naji}
\address{Kavli Institute for Theoretical Physics, University of California Santa Barbara, CA 93106, U.S.A.}
\address{Dept. of Physics, and Dept. of Chemistry and Biochemistry, 
University of California, Santa Barbara, CA 93106, USA\\}
\author{R. Podgornik}
\address{Kavli Institute for Theoretical Physics, University of California Santa Barbara, CA 93106, U.S.A.}
\address{Dept. of Physics, Faculty of Mathematics and Physics, University of 
Ljubljana and Dept. of Theoretical Physics, J. Stefan Institute, SI-1000 
Ljubljana, Slovenia}
\address{ Laborarory of Physical and Structural Biology, National 
Institutes of Health, MD 20892, USA}

\pacs{05.40.-a, 77.22.-d}

\begin{abstract} 
We study the thermal Casimir effect between two thick slabs composed of plane-parallel 
layers of random dielectric materials interacting across an intervening homogeneous dielectric. It is 
found that the effective interaction at long distances is self averaging and is given by a description 
in terms of effective dielectric functions. The behavior at short distances becomes random (sample 
dependent) and is dominated by the local values of the dielectric function proximal to each other across 
the dielectrically homogeneous slab. 
\end{abstract}

\maketitle

Systems with spatially varying dielectric functions exhibit effective van der Waals interactions arising 
from the interaction between fluctuating dipoles in the system \cite{mah1976,par2006}. These fluctuation 
interactions have two distinct components: (i) a classical or thermal component due to the zero 
frequency response of the dipoles and (ii) a quantum component due to the non zero frequency/quantum 
response of the dipoles. Despite the clear physical differences in these contributions, the mathematical 
computation of the corresponding interaction is almost identical and boils down to the computation of an 
appropriate functional determinant. The full theory taking into account both of these component 
interactions is the celebrated Lifshitz theory of van der Waals interactions \cite{dzy1961}, based on 
boundary conditions imposed on the electromagnetic field at the bounding surfaces and the 
fluctuation-dissipation theorem for the electromagnetic potential operators. The original Casimir 
interaction \cite{casimir} is obtained in the limit of zero temperature and ideally polarizable bounding
surfaces. At non-zero temperature the contribution of the zero frequency modes to the Lifshitz
theory yields the classical thermal Casimir effect which is due to the non-retarded van der Waals 
interactions.  
\begin{figure}
\includegraphics[scale=0.55]{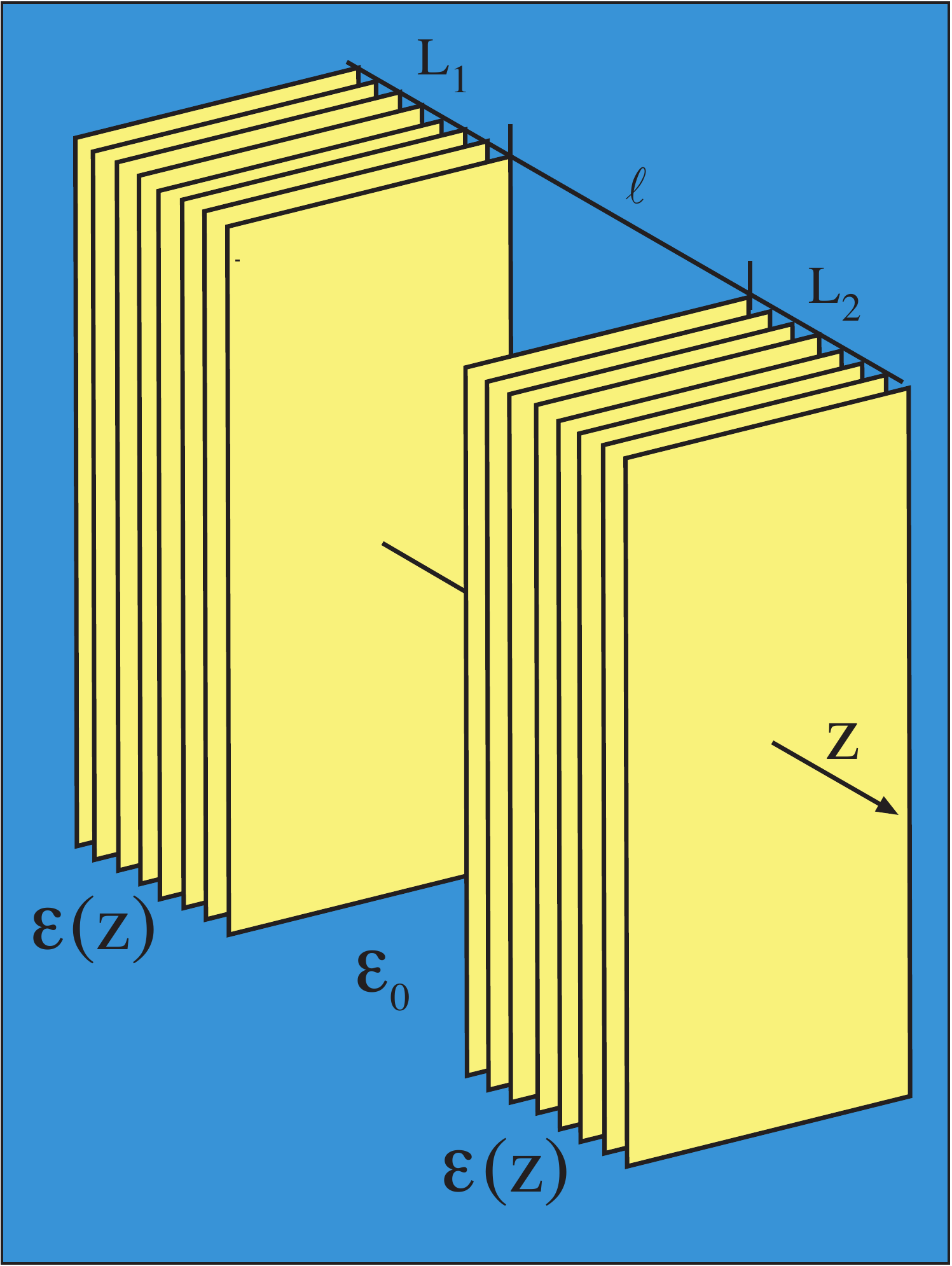}
\caption{A schematic presentation of the model. Two finite slabs, (1) and (2), with disordered 
plane-parallel dielectric layers interacting across a dielectrically homogeneous slab of thickness 
$\ell$. $z$ axis is perpendicular to the plane of the slabs.}
\label{schematic}
\end{figure}
The major mathematical problems in the computation of Casimir type interactions (setting aside the 
experimental and theoretical challenges to determine the correct dielectric behavior) are (i) the 
application of the Lifshitz approach to non-trivial geometries and (ii) taking into account local 
inhomogeneities in the dielectric properties of the media, always present in realistic systems. In this 
paper we will address the latter.

We consider the thermal Casimir interaction for the case where the local dielectric function is a 
random variable in the transverse direction. Specifically we will consider the interaction between two 
thick parallel dielectric slabs, separated by a homogenous dielectric medium, see Fig. 
(\ref{schematic}). The thickness of both disordered dielectric slabs are $L_1$ and $L_2$ respectively 
and their separation is denoted by $\ell$. In what follows we will study the limit of infinite slabs 
{\em i.e.} $L_1,\ L_2\to \infty$. The dielectric response within the two slabs is constant in the planes 
perpendicular to the slab normal, but varies in the direction of the surface normal. It is well known 
that this problem can be solved in the case where the dielectric constants of the slabs do not vary 
\cite{par2006} and the result can be tentatively applied to the case of fluctuating dielectric functions 
{\sl via} an effective medium theory which consists of replacing the fluctuating dielectric functions by 
an effective (spatially constant within each of the slabs) dielectric tensor. The most commonly used 
approximation is that where the local dielectric tensor is replaced by the effective dielectric tensor 
\cite{mah1976,par2006}, {\em i.e.}
\begin{equation}
\epsilon_{ij}({\bf x}) \to \epsilon^{(e)}_{ij}, \label{eqdiap}
\end{equation}
with the bulk dielectric tensor defined via $\epsilon^{(e)}_{ij}\langle E_j\rangle = \langle 
\epsilon_{ij} E_j\rangle$. The use of the effective dielectric function is not easily justifiable 
mathematically as an approximation, although physically the effective dielectric function clearly does 
capture the bulk response to constant electric fields. We shall see that, for the random layered 
dielectric model studied here, the effective dielectric constant approximation of Eq. (\ref{eqdiap}) 
does in fact give the correct value of the thermal Casimir interaction when the two slabs are widely 
separated. This can be expected on physical grounds since the fluctuating electromagnetic field modes with 
smallest wave-vectors (corresponding to variations on large scales) dominate the Casimir interaction for 
large inter-slab separation.  The dielectric response of the material to a constant electric field is 
given by the effective dielectric constant and if the wave-vector dependent response is suitably 
analytic near ${\bf k}=0$ we expect that $\epsilon^{(e)}_{ij}({\bf k})\sim 
\epsilon^{(e)}_{ij}(0)=\epsilon^{(e)}_{ij}$ for $|{\bf k}|\ll 1$.

In this letter we introduce a path integral formalism to compute the thermal Casimir free energy between 
two semi-infinite dielectric slabs which are composed of layers with varying dielectric function. Our 
formulation allows us to show rigorously that for large inter-slab separations the leading order 
contribution to the interaction is self averaging and is equivalent to that obtained by replacing each 
slab with a homogeneous (though non-isotropic medium) with a dielectric tensor equal to the effective 
(bulk) dielectric tensor of the disordered medium. The short distance behavior of the interaction is 
random and we show, as would be expected on physical grounds, that it is dominated by the precise value 
of the dielectric constants at the two opposing slab faces.

The Hamiltonian  for the zero frequency fluctuations of the 
electrostatic field in a dielectric medium is given by the classical electromagnetic field energy
\begin{equation}
H[\phi]= {1\over 2}\int d{\bf x}\  \epsilon({\bf x}) \left(\nabla \phi({\bf x})\right)^2
\end{equation}
and the corresponding partition function is given by the functional integral $Z = \int d[\phi] 
\exp(-\beta H[\phi])$. Differences in dielectric functions lead to the thermal Casimir effect. Here we 
will consider layered systems where the dielectric function $\epsilon$ depends only on the $z$ direction 
$\epsilon({\bf x}) = \epsilon(z)$. If we express the field $\phi$ in terms of its Fourier modes in the 
plane perpendicular to $z$, and we take the area perpendicular to $z$ as $A$, with wave-vector ${\bf k} 
= (k_x, k_y)$, then the Hamiltonian can be written as $H = \sum_{\bf k} H_{\bf k}$ with
\begin{equation}
H_{\bf k} = {1\over 2} \int dz \epsilon(z) \left( \left|{d{\tilde \phi}(z,{\bf k})\over dz} \right|^2
+ {\bf k}^2 |{\tilde \phi}(z,{\bf k})|^2\right).
\end{equation}
A direct consequence of this decomposition is that the partition function 
can be expressed as a sum over the partition functions of individual modes $Z_{\bf k}$ as
$\ln(Z) = \sum_{k} \ln(Z_{\bf k})$
where
\begin{equation}
Z_{\bf k} = \int d[X] \exp\left(-{1\over 2} \int dz\  \epsilon(z) \left[\left({dX\over dz}\right)^2 
+ k^2 X^2\right]
\right).
\end{equation}
Here $k=|{\bf k}|$ and we have taken into account that the field $\phi$ is real. 

The problem of computing the interaction between slabs composed of layers of 
finite thickness can be studied using a transfer matrix method 
\cite{pod2004}. However we will use a method based on the Feynman path integral 
which is particularly well suited to the study of systems where the 
dielectric function can vary continuously \cite{pvv}. If we specify the starting 
and ending points of the above path integral, we see that it has to be of a harmonic 
oscillator form  defined by
\begin{equation}
K(x,y;z) = \int d[X] \exp\left(-{1\over 2} \int dz M(z) \left[\left({dX\over dz}\right)^2 + 
\omega^2 X^2\right]
\right),
\end{equation}
which can be computed using the generalized Pauli - van Vleck formula \cite{pvv,inprep} telling us that $K$ must have the general form
\begin{equation}
K(x,y;z) = \left( {b\over 2\pi}\right)^{1\over 2}\exp\left(-{1\over 2} a_i(z)x^2 -{1\over 2} a_f(z)y^2 +b(z) xy\right).
\label{eq:kernel}
\end{equation}
We may now write down evolution equations for the coefficients $a_i,\ a_f$ 
and $b$ using the Markovian property of the path integral, $K(x,y;z+z')= 
\int dw~ K(x,w;z) K(w,y;z')$  \cite{pvv,inprep}, which can also be used to
prove the generalized Pauli - van Vleck formula. We obtain the evolution equations
\begin{eqnarray}
{da_i\over dz} &=& -{b^2\over M},\ \ \  {db\over dz} = -{ba_f\over M}, \noindent \\
{da_f\over dz} &=& M\omega^2 - {a_f^2\over M}\;. \label{eqaf}
\end{eqnarray}
We thus find that the $\ell$-dependent part of the free energy of the mode 
$\bf k$ (up to a bulk term which can be subtracted off to get the interaction energy) 
is given by
\begin{equation}
F_{\bf k} = {k_B T\over 2} \ln\left( 1- {(a_f^{(1)}(k)-\epsilon_0k) (a_f^{(2)}(k)-\epsilon_0 k) \over
(a_f^{(1)}(k)+\epsilon_0k)(a_f^{(2)}(k)+\epsilon_0k)} e^{-2k\ell }\right)\;,
\label{freq}
\end{equation}
and the total $\ell$ dependent free energy is $F = \sum_{\bf k} F_{\bf k}$. Here $a^{(1,2)}_f(k)$ are the
solutions to Eq. (\ref{eqaf}) evaluated at the opposing faces of each slab (1) and (2) respectively.

In order to evaluate the integrals of $a^{(1,2)}_f(k)$, one first has to solve equations of motion Eqs. 
(\ref{eqaf}) to get the $z$ dependence of $a_f(k, z)$ and then proceed to the integrals that enter Eq. 
(\ref{freq}). The evolution equation for $a_f(k)$ for either slab can be read off from Eq. (\ref{eqaf}) 
and is given by
\begin{equation}
{da_f(k,z)\over dz} =\epsilon(z) k^2 - {a_f^2\over \epsilon(z)}.
\label{equndef}
\end{equation}

An appropriate Hopf-Cole transformation \cite{inprep} shows this formalism to be equivalent to the 
transfer matrix method \cite{pod2004} or to the density functional method \cite{veble} for evaluating 
the van der Waals forces. This nonlinear formulation of an essentially linear problem simplifies 
the analysis of the effect of disorder in a similar way as it does in quantum problems \cite{itzykson}. 
We now write $a_f^{(i)}(k, z) = k\alpha^{(i)}(k, z)$ and if the distributions of the $\alpha^{(i)}(k. 
z) = y$ are given by $p_i(k,y)$ we find that, in three dimensions the average of the $\ell$ 
dependent free energy is given by
\begin{eqnarray}
\langle F \rangle &=& {k_B TA\over 4\pi }\int dk k \int dy_1\int dy_2 p_1(k,y_1)p_2(k,y_2)\nonumber \\
&&\ln\left( 1- {(y_1 - \epsilon_0) (y_2 - \epsilon_0 ) \over
(y_1 +\epsilon_0)(y_2 + \epsilon_0)} e^{-2k\ell}\right),\label{free1}
\end{eqnarray}
where the angled bracket on the l.h.s. indicates the disorder average over the dielectric 
function within the slabs and we have assumed that the realizations of the 
disorder in the two slabs are independent. 

Let us first investigate the form of van der Waals interaction free energy in the limit of large 
separations between the two slabs. The equation obeyed by $\alpha$ can be written as
\begin{equation}
{d\alpha(k,\zeta)\over d\zeta} =\epsilon({\zeta/ k})  - {\alpha^2\over \epsilon({\zeta/ k})},\label{eqmal}
\end{equation}
with $\zeta = zk$. When $k$ is small $\epsilon({\zeta/ k})$ varies very 
rapidly and so becomes decorrelated from the value of $\alpha$.  The Laplace 
transform for the probability density function of $\alpha$ is defined by
${\tilde p}(k,s,\zeta) = \int_0^\infty dy \exp(-sy) p(k,y,\zeta) = \langle \exp(-s \alpha(k,\zeta))\rangle\;
$ and, from the equation of motion Eq. (\ref{eqmal}), obeys
\begin{eqnarray}
&&-{1\over s} {d{\tilde p}(k,s,\zeta) \over d\zeta}=\nonumber \\ && 
\left\langle \epsilon({\zeta/ k})  \exp(-s \alpha(k,\zeta))  - 
{\alpha^2\over \epsilon({\zeta/ k})} \exp(-s \alpha(k,\zeta))\right\rangle\;. \nonumber \\
\end{eqnarray}
Assuming that $k$ is small and thus that $ \alpha(k,\zeta)$ and $\epsilon({\zeta/ k})$ 
are decorrelated, we can write
\begin{equation}
-{1\over s} {d{\tilde p}(k,s,\zeta) \over d\zeta} = \nonumber \\
\langle \epsilon\rangle {\tilde p}(k,s,\zeta)
-\left\langle {1/ \epsilon}\right\rangle {d^2\over ds^2}{\tilde p}(k,s,\zeta)\;.
\end{equation}
As we are interested in the limit of thick slabs  it suffices to know the equilibrium distribution of
this equation which is given by   
$\lim_{\zeta\to \infty}{\tilde p}(k,s,\zeta)= \exp(-\epsilon^* s)$
with 
\begin{equation}
\epsilon^* = \sqrt{\frac{\langle\epsilon\rangle}{\left\langle {1/ \epsilon}\right\rangle}}\;.  \label{eqstar1}
\end{equation}
Inverting the Laplace transform then gives the equilibrium distribution $p(y,k)  = \delta(y- \epsilon^*) 
$ at small $k$.  When $\ell$ is large the integral in Eq. (\ref{free1}) is dominated by the small $k$ 
behavior and we may use the analysis presented above, to give the following asymptotic form for the 
interaction free energy
\begin{equation}
\langle F \rangle(\ell \to \infty) \sim {k_B TA\over 16\pi \ell^2 }
\int udu \ln\left( 1- \Delta_1^*\Delta_2^* e^{-u}\right) = -{H^*A\over \ell^2}\;,
\label{free2}
\end{equation}

with $\Delta_i^* = {(\epsilon^*_i - \epsilon_0)/(\epsilon_i^* +\epsilon_0)}$ and where $\epsilon_i^*$ 
are defined via Eq. (\ref{eqstar1}). The subscript $i$ on the angled brackets signifies that we are 
averaging the dielectric function in the slab $i$.  The term $H^*$ defines an effective 
disorder-dependent Hamaker coefficient. This therefore justifies physical arguments replacing the random 
layered material by an effective anisotropic medium where the dielectric tensor is has the form 
$\epsilon^{(e)}_{zz} =\epsilon_{||}$ and $\epsilon^{(e)}_{xx}=\epsilon^{(e)}_{yy} = \epsilon_{\perp}$, 
all other terms being zero by symmetry. The term $\epsilon_{||}$ is the effective dielectric function in 
the $z$ direction $\epsilon^{(e)}_{||} = 1/\left\langle 1/\epsilon\right\rangle$, and the perpendicular 
components are given by $\epsilon^{(e)}_{\perp} = \langle \epsilon \rangle$. The expressions for 
$\epsilon^{(e)}_{||}$ and $\epsilon^{(e)}_{\perp}$ follow simply from the fact that in the perpendicular 
direction the dielectric function is obtained by analogy to capacitors in series and in the parallel 
direction by analogy to capacitors in parallel arrangement \cite{podg2}. The effective value, 
$\epsilon^*$, for dielectric constant of this system coincides with that of Eq. (\ref{eqstar1}) above 
\cite{inprep}.  This result shows that for large separations (where $\ell$ is much larger than the 
correlation length of the dielectric disorder)  the thermal Casimir interaction free energy is self 
averaging and agrees with that given by physical reasoning.

One would imagine that as the distance between the slabs is reduced, the 
result will be increasingly dominated by the slab composition at the two 
opposite faces \cite{par2006}. Indeed in the small $\ell$ limit Eq. (\ref{free1}) is dominated by 
the large $k$ behavior. The asymptotic behavior can be extracted if one assumes the {\sl ansatz} 
\begin{equation}
\alpha(z,k) = \sum_{n=0}^\infty {\alpha_n(z)\over k^n}. 
\end{equation}
Substituting this into Eq. (\ref{eqmal}) gives the following chain of equations for $\alpha_n(z)$
\begin{equation}
{1\over k}\sum_{n=0}^\infty {\frac{1}{k^n}\frac{d\alpha_n(z)}{dz}} = 
\epsilon(z) - {1\over \epsilon(z)}\sum_{n,m=0}^\infty
{\alpha_n(z)\alpha_m(z)\over k^{m+n}}\;.
\end{equation}
From here it is easy to see that to order $O(1)$ the leading asymptotic result of Eq. (\ref{ask}) is given by
\begin{equation}  
\alpha_0(z) = \epsilon(z)\;.
\end{equation}
The equation for the corrections ($n\geq1$) to this asymptotic limit  is
\begin{equation}
{d\alpha_{n-1}(z)\over dz} = -{1\over \epsilon(z)}\sum_{m=0}^n \alpha_m(z) \alpha_{n-m}(z),
\end{equation}
and the next two terms from this expansion yield 
\begin{eqnarray}
\alpha_1(z)  &=&-{1\over 2} {d\epsilon(z)\over dz}\;,\\
\alpha_2(z) &=&  {1\over 4}{d^2\epsilon(z)\over dz^2} - 
\frac{1}{8\epsilon(z)}\left({d\epsilon(z)\over dz}\right)^2\;.
\end{eqnarray}
It is straightforward to realize that these terms generate $O(1/\ell)$ corrections to the asymptotic result
which are subdominant when $\ell$ is large. Thus to the leading order
\begin{equation}
\alpha(z,k) \approx \alpha_0(z) = \epsilon(z) 
\label{ask}
\end{equation}
and from here it follows straightforwardly that
\begin{equation}
\lim_{k\to \infty} p_i(y,k) = \rho_i(\epsilon)\label{largek}
\end{equation}
where $\rho_{i}$ is the probability density function of $\epsilon(z)$ in medium 
$i$. This result is easily understood from the physical discussion above.  The average of the thermal Casimir interaction free energy Eq. (\ref{free1}) in the small separation limit is then given by
\begin{eqnarray}
\langle F\rangle (\ell \to 0) &\sim& {k_B TA\over 16\pi \ell^2 }\int udu\int \rho_1(\epsilon_1)\rho_2(\epsilon_2) 
d\epsilon_1 d\epsilon_2 \nonumber \\&&\ln\left( 1- \Delta_1\Delta_2 e^{-u}\right),
\label{freesmall}
\end{eqnarray}
with $\Delta_i = {(\epsilon_i-\epsilon_0)/( \epsilon_i+\epsilon_0)}$.
The forms of the thermal Casimir interaction free energy are thus given by Eqs. (\ref{freesmall}) and (\ref{free2}) in the small and large interslab separation limits respectively. 

In the limit of large separation between the slabs we have obtained the limiting behavior of the thermal 
Casimir effect and shown that the free energy is given by self-averaging and that the 
distributions of $\alpha(k,z)$ are strongly peaked. It can be shown \cite{inprep} that the attraction
at large separation between two (statistically identical)  homogeneous media (with $\epsilon =\langle 
\epsilon\rangle$) is stronger than that between the two fluctuating media if $\langle1 
/\epsilon\rangle^{-1} > \epsilon_0$. However it is always weaker if $\langle \epsilon\rangle 
<\epsilon_0$.  So, depending on the details of the distribution of the fluctuating 
dielectric response in the two slabs and the dielectric response of the medium in-between, the effective 
interaction at large inter-slab separations can be stronger or weaker than that for a uniform medium 
with a dielectric constant equal to the mean dielectric function of the fluctuating media.

For small separations the interaction free energy is a random variable which has to be averaged over the 
probability density function of the dielectric functions in the media composing the two interacting 
slabs. The intermediate length scales can be analyzed {\sl via} perturbation theory
\cite{inprep}, and there may also exist models of disorder that can be treated exactly. The nonlinear 
formulation of the problem presented here should be equally useful to treat the case of 
deterministically varying dielectric functions and could open up a useful computational framework for 
designing materials where the effective interaction can be tuned to induce attractive or repulsive 
forces depending on the separation, for practical applications \cite{apps} .

This research was supported in part by the National Science Foundation under Grant No.~PHY05-51164. 
D.S.D acknowledges support from the Institut Universtaire de France. R.P. would like to acknowledge the 
financial support by the Agency for Research and Development of Slovenia, Grants No. P1-0055C, No. 
Z1-7171, No. L2- 7080. This study was supported in part by the Intramural Research Program of the NIH, 
National Institute of Child Health and Human Development.

\end{document}